\documentclass[%
 reprint,
superscriptaddress,
 amsmath,amssymb,
 aps,
]{revtex4-1}

\usepackage[dvipsnames]{xcolor}
\usepackage{graphicx}
\usepackage{dcolumn}
\usepackage{bm}
\usepackage{amsthm}
\usepackage{amscd}
\usepackage{cancel}
\usepackage{dsfont}
\usepackage{hyperref}
\usepackage{ulem}
\usepackage{multirow}

\hypersetup{
	colorlinks = true,
	urlcolor   = blue,
	linkcolor  = blue,
	citecolor  = blue
}

\begin{document}


\title{Robust architecture for programmable universal unitaries}

\author{M. Yu. Saygin}
\email{saygin@physics.msu.ru}
\affiliation{Quantum Technologies Center, Faculty of Physics, Lomonosov Moscow State University, Leninskie gory 1, building 35, 119991, Moscow, Russia}%

\author{I. V. Kondratyev}
\affiliation{Quantum Technologies Center, Faculty of Physics, Lomonosov Moscow State University, Leninskie gory 1, building 35, 119991, Moscow, Russia}%
 
\author{I. V. Dyakonov}%
\affiliation{Quantum Technologies Center, Faculty of Physics, Lomonosov Moscow State University, Leninskie gory 1, building 35, 119991, Moscow, Russia}%

\author{S. A. Mironov}
\affiliation{Institute for Nuclear Research of the Russian Academy of Sciences,
60th October Anniversary Prospect, 7a, 117312 Moscow, Russia}%
\affiliation{Institute for Theoretical and Experimental Physics,
Bolshaya Cheriomyshkinskaya, 25, 117218 Moscow, Russia}%
\affiliation{Moscow Institute of Physics and Technology, Institutski pereulok, 9, 141701, Dolgoprudny, Russia}%

\author{S. S. Straupe}
\affiliation{Quantum Technologies Center, Faculty of Physics, Lomonosov Moscow State University, Leninskie gory 1, building 35, 119991, Moscow, Russia}%

\author{S. P. Kulik}
\affiliation{Quantum Technologies Center, Faculty of Physics, Lomonosov Moscow State University, Leninskie gory 1, building 35, 119991, Moscow, Russia}%

\date{\today}

\begin{abstract}

The decomposition of large unitary matrices into smaller ones is important, because it provides ways to realization of classical and quantum information processing schemes. Today, most of the methods use planar meshes of tunable two-channel blocks, however, the schemes turn out to be sensitive to fabrication errors. We study a novel decomposition method based on multi-channel blocks. We have shown that the scheme is universal even when the block`s transfer matrices are chosen at random, making it virtually insensitive to errors. Moreover, the placement of the variable elements can be arbitrary, so that the scheme is not bound to specific topologies. Our method can be beneficial for large-scale implementations of unitary transformations by techniques, which are not of wide proliferation today or yet to be developed.

\end{abstract}

\maketitle

Linear transformations of multiple channels are ubiquitous in many applied fields, as well as in fundamental research.
The implementations of linear multi-channel transformations are particularly important in the  optical context, in which they 
are used  as indispensable components of communication and computing systems, in particular, those based on novel approaches to information processing. For example, universal optical interferometers~-- devices  capable of performing arbitrary linear  transformations~--  are exploited as mode unscramblers~\cite{Unscrambler2017} or as a part of photonic neural networks~\cite{Shen2017,Hughes2018}.

In addition to classical applications, universal optical interferometers  play an important role in the implementations of quantum algorithms. In particular, multi-channel interferometers are a necessary part of the  promising quantum computing platform that leverages linear-optical circuits and non-classical properties of photons to realize quantum algorithms \cite{Harris2018, Rudolph2017}. Recent works have demonstrated the versatility of linear-optical quantum systems and their ability to perform a number of quantum computing tasks ranging from the well-known algorithms \cite{Politi2009, Zhou2013} to the more specific ones, such as boson sampling \cite{Spring2012, Broome2012, Crespi2013} and variational algorithms \cite{Peruzzo2014, Wiebe2016, Wang2017}. The efficient realization of the later two classes of algorithms was made possible by the universal programmable  interferometers \cite{Carolan2015}. 

The complexity of the multi-channel scheme, which is quantified by the number of channels in the interferometer and its programmability, defines its practical utility. These characteristics, however, are largely determined by a particular architecture used to construct the interferometer. Nowadays, the most widely used universal architectures belong to different planar varieties. One reason for this is the fact that planar schemes are easier to realize by the mature integrated photonics technology. On the other hand, methods have been devised that enable decomposition of large unitary matrices into planar meshes of smaller building blocks. 
 
The decomposition methods of wide use today are obtained from the unitary matrix factorization theorem proposed in \cite{Hurwitz1897}, which was adopted to construct planar reconfigurable schemes, for example, in  \cite{Reck1994, Clements2016}.  In the linear-optical context such a multi-channel scheme has a form of a mesh, constructed out of tunable Mach-Zehnder interferometers (MZIs), thereby enabling the whole interferometer to be reconfigured by tuning the variable phase elements~\cite{Reck1994,Clements2016}. For these schemes to be universal, the beam-splitters that constitute the MZIs should necessarily be balanced, therefore, imposing strict requirements on the fabrication tolerances. However, in practical implementations this condition is not fully satisfied, so that the multi-channel transformation is universal only to a certain degree~\cite{Burgwal2017}. This negative effect becomes more pronounced as the interferometer scheme is scaled up. Therefore, the realization of sophisticated information processing algorithms with optics calls for the development of novel architectures for multi-channel interferometers that are more resilient to the implementation errors. 

In this work we explore an alternative approach to bulding large-scale univeral interferometers, which is drastically different from the known methods based on meshes of two-channel blocks. It is based on sequences of fixed multi-channel mixing blocks instead of arrays of balanced beam-splitters. As a result, the interferometric scheme provides the freedom to choose the transfer matrix of the blocks out of the continuous unitary space. Surprisingly, such schemes turn out to be extremely robust to perturbations in the transfer matrices of the blocks, in fact, as we show below, they may be chosen at random, and the desired transformation may be dialed later on the manufactured device by tuning the phaseshifts only. In addition, the placement of the variable phase shifts in the scheme can also be chosen arbitrary. Since no relation to the scheme topology is implied in the analysis, a wide variety of technologies can be used to realize it in practice, for example, those based on frequency and temporal encoding, that may potentially enlarge the scale of the universal interferometric schemes.

\textit{Methods of decomposition of unitary matrices.} ---
Several types of parametrization for unitary matrices are known~\cite{Hurwitz1897,Jarlskog2005,Dita1982,Ivanov2008}, 
however, only few of them give a recipe to 
build up large devices out of elementary blocks that a convenient to realize in practice. Linear optics is the most representative domain of practical application for multi-channel transformations, where their implementations are employed to transform the input vector of field amplitudes $\mathbf{a}^{(in)}$ into the output vector $\mathbf{a}^{(out)}$ by transformation, described by a matrix $U$: $\mathbf{a}^{(out)} = U\mathbf{a}^{(in)}$. Following the decomposition methods, the required  matrix $U$, that can be regarded as a $\mathrm{SU}(N)$ transformation, is obtained by setting appropriate values of some parameters, characterizing the elements in the actual physical network \cite{Reck1994, Clements2016}. 

A natural building block for multi-channel schemes in the context of optics is a MZI consisting of a pair of static balanced beam-splitters and two variable phase shifters. Adding a third phase to the MZI allows to cover an entire $\mathrm{SU}(2)$ group \cite{Tilma2002} with the $2\times2$ transformation matrix:
    \begin{equation}\label{eq::su2_generation}
       e^{i\psi}U_{\text{MZI}}(\theta,\varphi)=e^{i(\psi-\pi)/2X}e^{i(\theta+\pi)Z}e^{i(\varphi-\pi)/2X},
    \end{equation}
where $X$ and $Z$ are the Pauli matrices, $\varphi$, $\theta$ and $\psi$ are the phase parameters.

Although, in principle any fixed beamsplitter transformation mixing a pair of optical modes densely generates $\mathrm{SU}(N)$ \cite{Bouland2014}, a more efficient way is to implement a network scheme allowing to program the required unitary using tunable MZIs.
The experiments reported in \cite{Carolan2015, Harris2017, Harris2018, Dyakonov2018} have demonstrated the capabilities of such a network design to experimentally approximate arbitrary unitaries. However, the performance of the linear optical network heavily depends on the quality of individual MZI elements. 
For example, the work \cite{Burgwal2017} studied the effect of individual beam-splitter errors on the overall fidelity and concluded that the beam-splitter reflectivity errors of few percents diminish the quality of the unitary transformations significantly. This makes a subset of unitary transformations unavailable for the programmable linear optical network composed of imperfect optical elements \cite{Russell2017}. 

\begin{figure}
\includegraphics[width=0.9\linewidth]{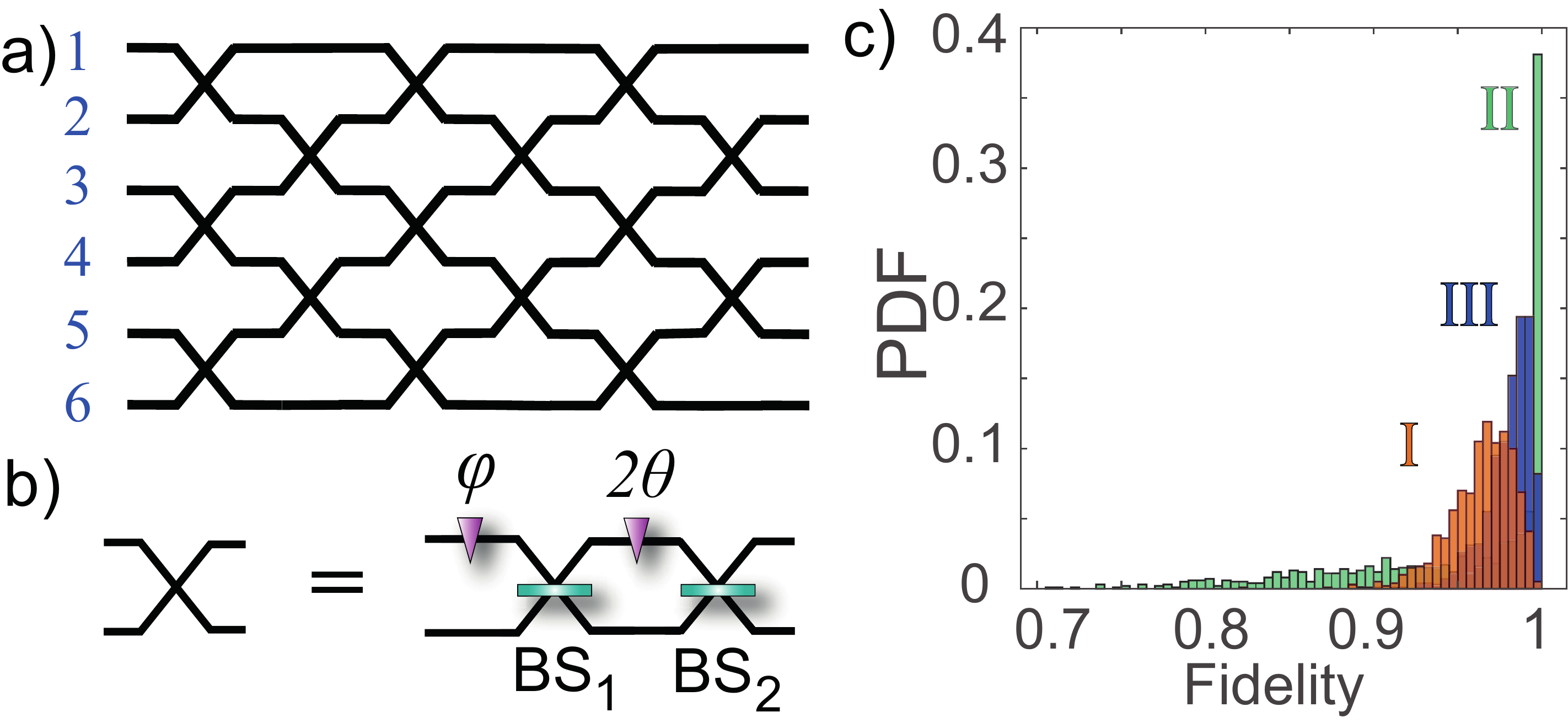}%
\caption{\label{fig:clements_imbalances} 
The planar universal design of a multi-channel decomposition proposed in \cite{Clements2016}.
(a) The  scheme layout  for $N=6$.
(b) The building block of the scheme -- an MZI with two tunable phase shifts, $\theta$ and $\varphi$, and static balanced beam-splitters BS$_1$ and BS$_2$.
(c) The histograms illustrating fidelity of transformations (see formula \eqref{eqn::fidelity}) for multi-channel schemes with $N=10$, corresponding to three error models: 
(I) all static beam-splitters are biased by an equal angle $\alpha$, chosen randomly in the range  $[0;20]$ degrees (red); 
(II) the beam-splitters are biased independently by angles $\alpha_j$  distributed in the range  $[-20;20]$  degrees (green);
(III) the beam-splitters are biased independently by angles $\alpha_j$  distributed in the range  $[0;20]$  degrees (blue).
Each histogram was obtained by an optimization procedure that found a global maximum of fidelity with respect to the set of phase shifts, collected over $1000$ randomly generated unitary matrices.}
\end{figure}

As an illustrative example, consider the   decomposition proposed in \cite{Clements2016}, which is  widely used today to construct  reconfigurable planar optical schemes. The corresponding scheme layout  is illustrated in Fig.~\ref{fig:clements_imbalances}a for a number of inputs and outputs $N=6$. Its main building block -- an MZI is depicted in Fig.\ref{fig:clements_imbalances}b. Fig.~\ref{fig:clements_imbalances}c illustrates the negative effect of the beam-splitter imbalances on the quality of the transformation, clearly witnessing the sensitivity of this decomposition to errors.

\begin{figure}
    \includegraphics[width=0.9\linewidth]{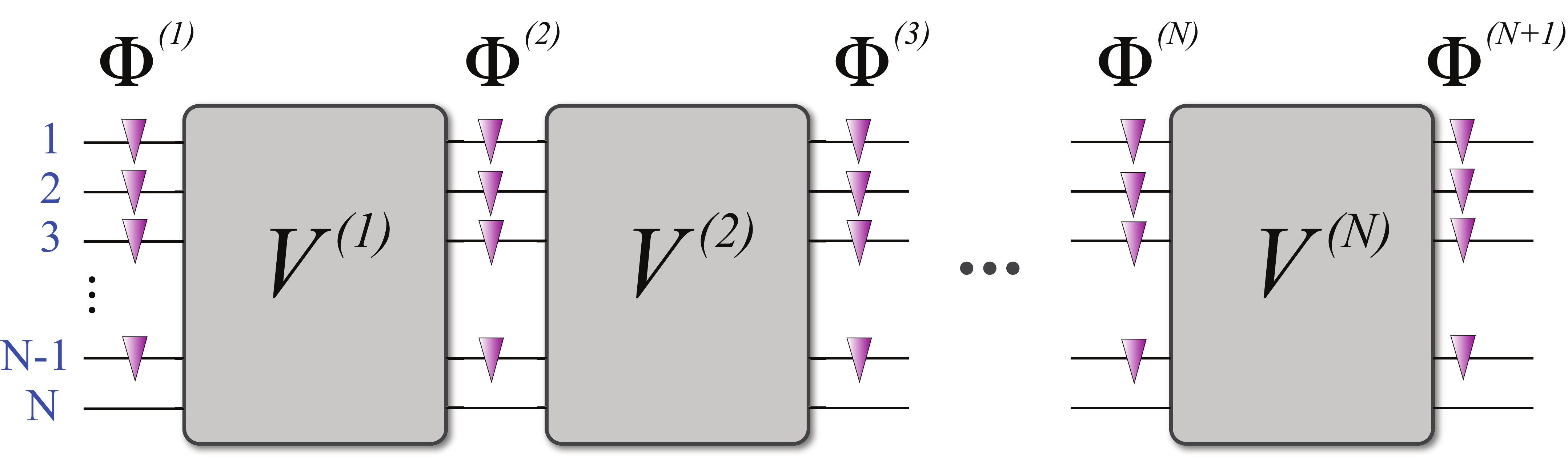}%
    \caption{\label{fig:general_scheme}  A schematic of a matrix decomposition~\eqref{eqn:OurDecomposition} the number of layers $K=N$ corresponding to $N-1$ phase shifts in each layer.
    }
\end{figure}

\textit{Layered decompositions with static multi-channel blocks.} ---
The decomposition we study in this work is based on static multi-channel blocks, rather than two-channel balanced beam-splitters. The schematic  of the decomposition is shown in Fig.~\ref{fig:general_scheme}. It is built up of multiple layers, that come in two types, which are stacked alternately. The variable layer consists of independent single phase shifts $\varphi_j$,
so that its $N$-channel transfer matrix has the diagonal form: 
$    \Phi(\vec{\varphi})=\text{diag}\left(\exp(i\vec{\varphi})\right),$
where $\vec{\varphi}=\left(\varphi_1,\varphi_2,\ldots,\varphi_N\right)$ is a vector of phase shifts, thus, we call it the phase layer. The other type of layer, in the following referred to as the mixing layer, introduces interaction between the channels that is required for multi-channel interference. It is the aim of our work to find out what transfer matrices $V^{(m)}$ should these mixing layers have in order for the whole scheme to be universal.
Moreover, we are interested in a general case, when the mixing layers can have different transfer matrices $V^{(m)}$, so that the overall transformation has the following form:
    \begin{equation}\label{eqn:OurDecomposition}
        U=\Phi^{(K+1)}V^{(K)}\Phi^{(K)}\cdot\ldots\cdot{}V^{(1)}\Phi^{(1)}
    \end{equation}
where $\Phi^{(m)}=\Phi(\vec{\varphi}^{(m)})$ with $\vec{\varphi}^{(m)}$ being the set of phases describing the layer with index $m$, $K$ is the number of mixing layers.  The  decomposition \eqref{eqn:OurDecomposition} enables a variety of schemes, each having different number of layers $K$ at a given $N$. Here, we describe the case of $K=N$ \footnote{See Supplemental Material [url] for other possible variants of the interferometer schemes}.

Previous works considered special cases of decomposition \eqref{eqn:OurDecomposition} with specific choice of $V^{(m)}$. A design of the form \eqref{eqn:OurDecomposition} with the mixing layers described by the discrete Fourier transform matrix:
\begin{equation}\label{eqn::DFT}
    U^{(DFT)}_{mn}=\frac{1}{\sqrt{N}}\exp\left[i\frac{2\pi}{N}(m-1)(n-1)\right],
\end{equation}
$m,n=1\ldots{}N$, implemented in a planar integrated photonic circuit was studied in \cite{Tang2017}. Work \cite{Zhou2018} deals with the mixing layers implemented by regions of coupled  waveguides. The authors of these works provided some numerical evidence in favor of universality of this specific types of circuits.
In both these works, the transfer matrices of the mixing layers were essentially defined with a specific system in mind. Besides, the number of phase shifts per layer was $N-1$, which is a maximum possible number. Relevant ideas concerning decomposition~\eqref{eqn:OurDecomposition} can be found in \cite{Lu2018,Lu2019,Tang2017}.
In this work we show that the class of transfer matrices $V^{(m)}$ suitable for universal programmable schemes is not bounded to some specific instances, to the contrary, it occupies a large part of the $\mathrm{SU}(N)$ group volume.

We first notice that by simple counting of independent parameters necessary to define an arbitrary unitary matrix, a universal scheme performing an SU($N$) transformation should have at least $N^2-1$ tunable phase shifts. 
We will restrict our attention to the schemes using exactly this minimal set of phases. 

\begin{figure*}
\includegraphics[width=0.9\linewidth]{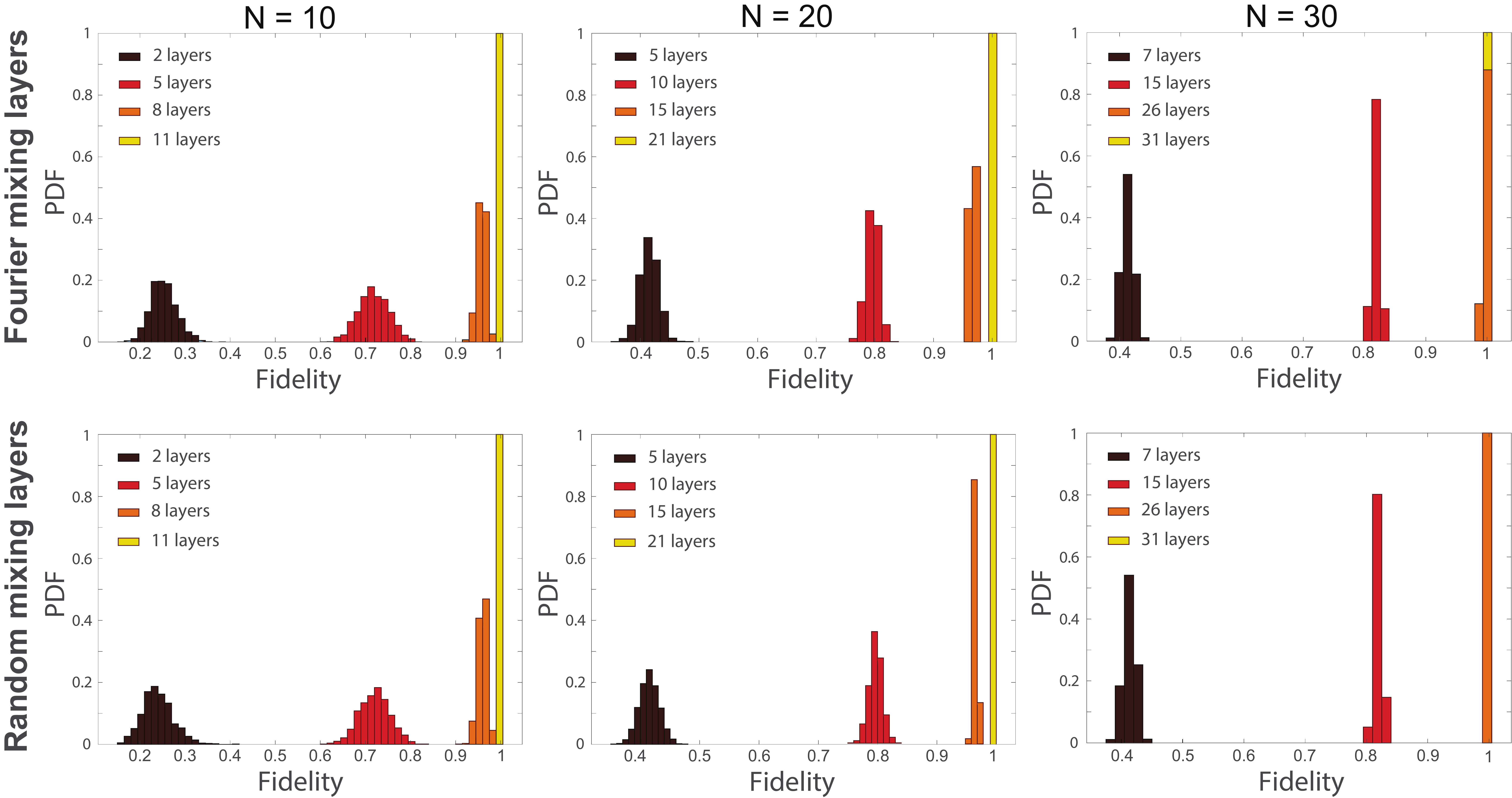}%
\caption{\label{fig:optimization_results} 
Fidelity of transformation for the decomposition depicted in Fig.~\ref{fig:general_scheme}a for $N=10,20$ and $30$ and different choice of the mixing layer transfer matrices.
The figures of the top row correspond to probability density function (PDF) at transfer matrices $V^{(m)}=U^{(DFT)}$, 
while the bottom row corresponds to all $V^{(m)}$ taken at random.
Each histogram is a collection of $1000$ fidelity values, corresponding to randomly sampled target unitary matrix. 
Different numbers of phase layers were used to illustrate the convergence of the schemes to universal, 
which correspond to the maximum phase layers of $N+1$.
}
\end{figure*}

We quantify performance of the decomposition \eqref{eqn:OurDecomposition} by fidelity, defined as:
\begin{equation}\label{eqn::fidelity}
    F(U,U_s)=\frac{1}{N^2}|\mathrm{Tr}\left(U^{\dagger}U_{s}\right)|^2,
\end{equation}
which compares a target unitary matrix $U_s$ and an actual transfer matrix realized by the decomposition $U$ \eqref{eqn:OurDecomposition}. Provided that the matrices $U$ and $U_s$ are equal up to a global phase, fidelity \eqref{eqn::fidelity} gets its maximum value $F=1$.

In contrast to the known decomposition of \cite{Clements2016}, which comes with an analytical procedure allowing to obtain a set of phases at which $F=1$ for any given $U_s$, we could not find an analytical solution for the general decomposition \eqref{eqn:OurDecomposition}. Therefore, a numerical optimization procedure has been used. Our numerical algorithm is based on the basinhopping algorithm  realized using the SciPy python library. Given a unitary matrix $U_s$, the algorithm was searching for a global minimum of infidelity $1-F$ over the space of phase vectors $\vec{\varphi}^{(m)}$ ($m=1\ldots{}K+1$). To decrease the chance of sticking into local minima, we used multiple runs of the basinhopping routine with random initial values of the phases. The algorithm has reasonable efficiency and has sub-exponential runtime dependence on the interferometer size \footnote{ See Supplemental Material [url] for the analysis of the efficiency of the numerical algorithm runtime and the effect of finite phase shift precision on transformation quality, which includes Refs~\cite{MillerPhysRevApplied}.}. This way, we have achieved the error of calculating the global minimum on the level of $\sim{}10^{-9}$ \footnote{The code is available on request}. Each numerical experiment involved  optimization over a series of $1000$  matrices $U_s$ drawn from the Haar random distribution using the method based on the QR-decomposition of random matrices from the Ginibre ensemble \cite{Mezzadri2006}. 

Let us now discuss the choice of the transfer matrices of the mixing layers, used in construction of multi-channel transformations. Surprisingly, not only specific fixed unitaries are suitable, but almost any unitary matrix will work as a mixing layer. Moreover, the mixing transformation need not be fixed, it can vary from layer to layer. To demonstrate that, we chose every matrix $V^{(m)}$ in \eqref{eqn:OurDecomposition} at random using the very same approach, described above for generation of $U_s$.
Fig.~\ref{fig:optimization_results} presents the results of optimization for decomposition depicted in Fig.~\ref{fig:general_scheme}a for the case of a fixed transfer matrix, namely a DFT matrix \eqref{eqn::DFT}, and random transfer matrices of the mixing layers. In the optimization process we varied the number of phase layers, ranging from small values containing less phase shifts than required for universality, which is done for illustrative purposes, to the proper number containing in total $N^2-1$ phase shifts. The worst value of infidelity corresponding to the latter case for all matrix sizes presented was not greater than $1-F<10^{-9}$ and its non-vanishing value is attributed to the finite accuracy of the implementation of the numerical algorithm. 

    \begin{figure}
        \includegraphics[width=0.9\linewidth]{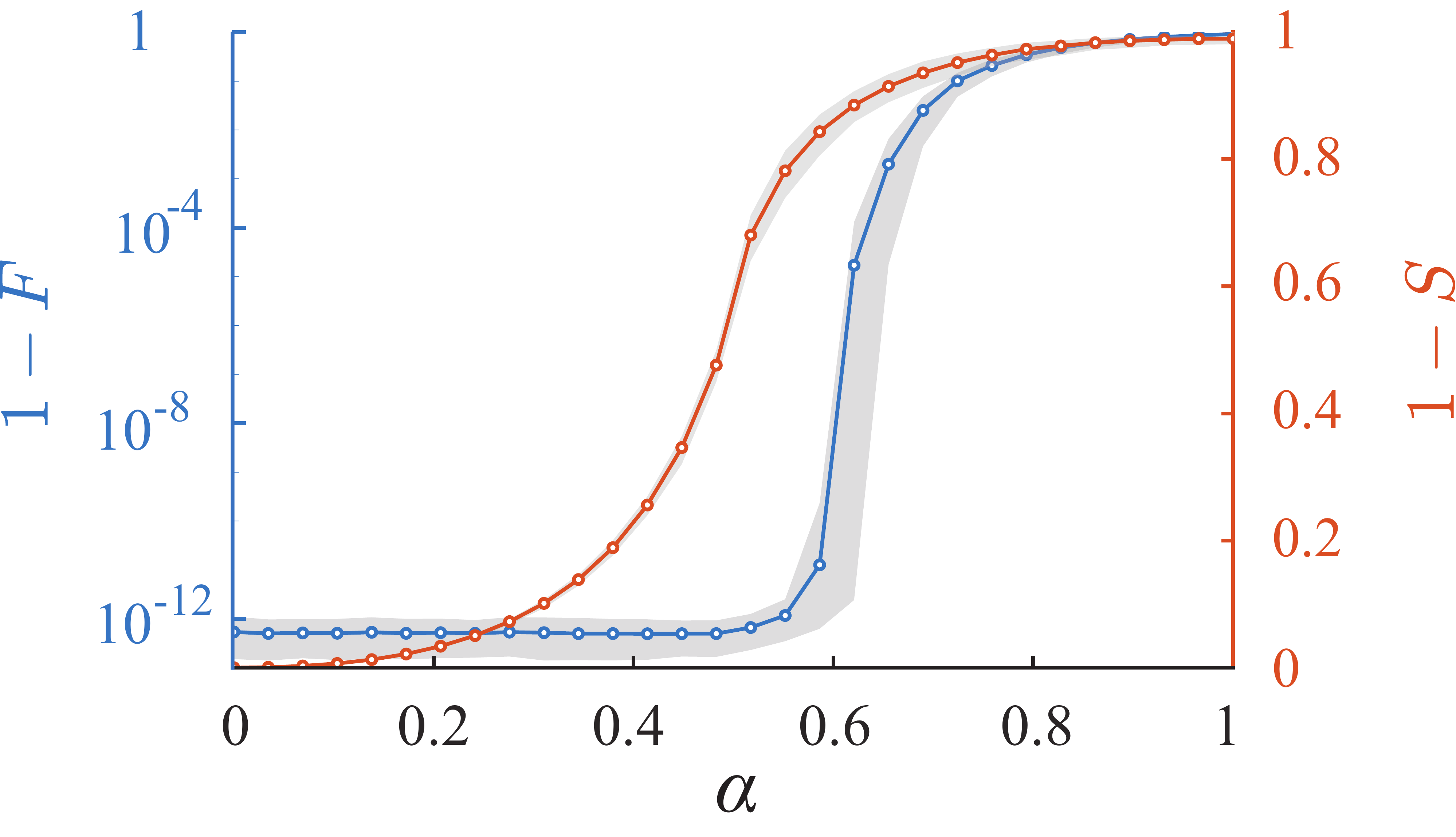}%
    \caption{\label{fig:robustness}
    Infidelity of transformation $1-F$ and dissimilarity of block transfer matrices $1-S$, where $S=\sum_{m=1}^NF(V_0^{(m)},V^{(m)}_{\alpha})/N$, as a function of parameter $\alpha$ at $N=10$. The circles correspond to the average values; the lower and upper boundaries of the shaded regions are the averages of the $10$ best and worst infidelities, respectively.
    }
    \end{figure}

To make sure, that our result is not a mere artifact of high-dimensional random matrix generation, we have considered a specific example of an SU(6) transformation corresponding to a probabilistic linear optical CNOT gate \cite{CNOT_Ralph}. We were able to reproduce this particular unitary with infidelity of $\sim{}10^{-8}$ using random mixing layers \footnote{ See Supplemental Material [url] for a particular example of linear-optical CNOT gate and robustness analysis of the interferometer architecture under consideration.}. 

To demonstrate that our architecture is resilient to practical constraints and errors that can undermine the universality of the interferometer, we have added random perturbations to the mode mixing layers according to the following procedure. First, we construct the parametrised matrix
   \begin{equation}\label{eqn:Talpha}
        T^{(m)}_{\alpha}=(1-\alpha)V^{(m)}_0+\alpha{}R,
    \end{equation}
which is a weighted sum of an initial matrix $V^{(m)}_0$, and a Haar-random perturbation $R$, modeling imperfections. Since $T^{(m)}_{\alpha}$ is non-unitary at $0<\alpha<1$, we then apply the singular value decomposition: $T^{(m)}_{\alpha}=W_{\alpha}^{(m,L)}D_{\alpha}^{(m)}W_{\alpha}^{(m,R)}$, where $W_{\alpha}^{(m,L)}$ and $W_{\alpha}^{(m,R)}$ are unitary matrices and $D_{\alpha}^{(m)}$ is a diagonal matrix. The unitary part of \eqref{eqn:Talpha} as: $V^{(m)}_{\alpha}=W_{\alpha}^{(m,L)}W_{\alpha}^{(m,R)}$. Even for a rather high value of $\alpha=1$ we have observed essentially no effect of the perturbation on the infidelity with an ideal CNOT matrix.

There are obvious examples of matrices which are not suitable for mixing layers, such as an identity matrix $I$ or permutation matrices. However, as our numerical results suggest, the relative volume of these matrices is negligible for large $N$. Parametrized families of matrices $V^{(m)}_{\alpha}$ allow us to analyze the worst case performance of our scheme. Changing $R$ in \eqref{eqn:Talpha} with $I$ we can see how close to a worst case we can get before the scheme becomes non-universal. For each value of $\alpha$ a series of $300$ target matrices $U_s$ and block transfer matrices $V^{(m)}_0$ were generated at random from a Haar-uniform distribution and fidelity \eqref{eqn::fidelity} was maximized over the space of phase shifts. Fig.~\ref{fig:robustness} demonstrates the infidelity of total transformation $1-F$ as a function of parameter $\alpha$ along with the infidelity for blocks $1-S$, where $S=\sum_{m=1}^NF(V_0^{(m)},V^{(m)}_{\alpha})/N$ is fidelity between the initial and the perturbed block transfer matrices. As can be seen from the figure, while $S$ for each block grows monotonically, the dependence of infidelity for the whole transformation has a clear threshold behaviour, namely, it stays on the constant level of accuracy of the numerical algorithm as far as $\alpha$ does not exceed value $\approx{}0.5$ after which it grows abruptly. We did not study the dependence of the threshold value of $\alpha$ on the interferometer size $N$ in details, but results for $N=5$ and $N=10$ suggest, that it may slightly increase with increasing $N$ \footnote{ See Supplemental Material [url] for the performance analysis of the interferometer architecture under consideration.}.

\textit{Theoretical framework.} ---
The strict proof of the universality implies showing that the  transformations $U(\{\vec{\varphi}^{(m)}\})$ should form a group under matrix multiplication and then prove that this group densely covers the $\mathrm{SU}(N)$ group of the corresponding dimension. We could not find the rigorous proof neither for the simplest nontrivial case of $N=3$ nor for the general case of arbitrary dimension. However, we have worked out preliminary considerations on the structure of the manifold described by \eqref{eqn:OurDecomposition} \footnote{ See Supplemental Material [url] for the analytical results, which includes
Refs.~\cite{5M,Kowalevicz:05,Coupled_mode_theory_Yariv_73,Spagnolo2012}}.

\textit{Conclusion.} ---
Our work demonstrates that multi-channel layered schemes provide an  alternative architecture for  universal linear-optical unitaries. This architecture is not limited to specific strictly predefined transfer matrices, such as, for example, a discrete Fourier transform as in previous proposals \cite{Huhtanen2007,Lu2018}. Namely, we have shown that the transfer matrices of the static blocks can be chosen from a continuous class of unitary matrices at random without 
sacrificing the quality of approximation for an arbitrary target unitary. This also comes immediately with resilience to errors that usually occur in implementations. In practice, the interferometer manufactured with imperfect mixing layers may be tuned to implement the desired transformation post-factum by tuning the phase-shifts only and optimizing the measured fidelity with the desired transformation \cite{Dyakonov2018}. Therefore, these results, showing that one does not necessarily need to carefully engineer the building blocks of the schemes to be able to reach almost any matrix in the unitary space, are of primary importance for experimental applications.

\textit{Acknowledgements}. This work was supported in part by RFBR grants \# 19-32-80020/19, 18-51-05015 and by the Foundation for the Advancement of Theoretical Physics and Mathematics (BASIS).  Authors are grateful to I.Bobrov, A.Mironov, A.Morozov, An.Morozov, M.Olshanetsky and Y.Zenkevich for fruitful discussions.

\bibliographystyle{apsrev4-1}
\bibliography{main}

\pagebreak
\onecolumngrid
\section*{Supplemental material}

\appendix
\section{Different variants of the universal interferometer schemes} \label{app::scheme_variants}

\begin{figure}[h!]
	\includegraphics[width=0.5\linewidth]{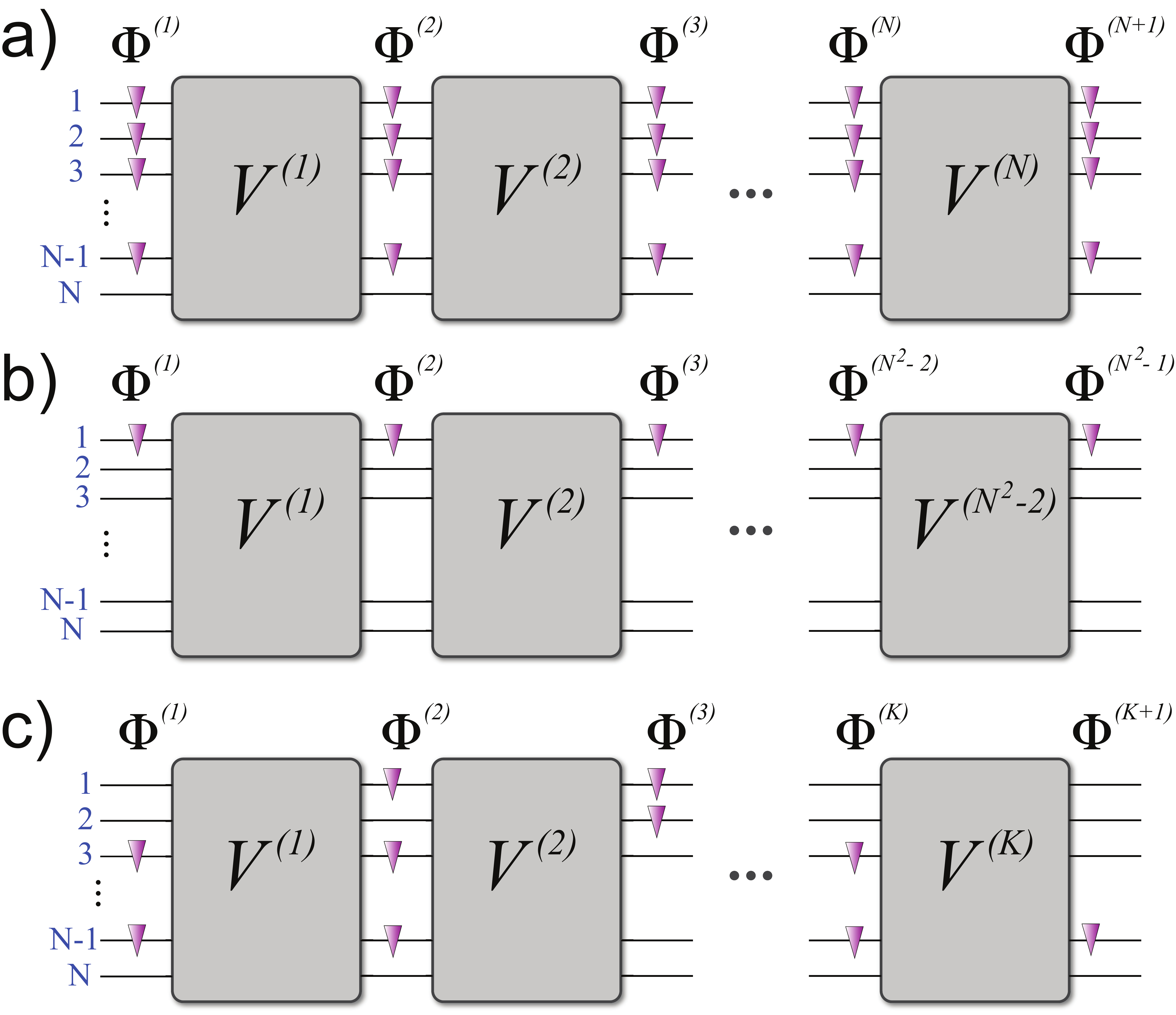}
	\caption{\label{fig:scheme_variants_App} The schematic of the possible varieties of matrix decomposition~\eqref{eq:OurDecompositionApp}, which differ in the number of phase shifts per layer and their placement:
		a) $N-1$ phase shifts ($K=N$); 
		b) $1$ phase shift ($K=N^2-2$); 
		c) from $2$ to $N-2$ phase shifts ($N+1\le{}K\le{}N^2-3$).}
\end{figure}

The general form of the layered decomposition 
\begin{equation}\label{eq:OurDecompositionApp}
U=\Phi^{(K+1)}V^{(K)}\Phi^{(K)}\cdot\ldots\cdot{}V^{(1)}\Phi^{(1)}
\end{equation}
may be realized in various schemes, each having different depths at a given $N$, as quantified by the number of mixing layers $K$. We categorize the schemes into three varieties, which are depicted in Fig.~\ref{fig:scheme_variants_App}.
The one shown in Fig.~\ref{fig:scheme_variants_App}a uses a maximum possible number of phase shifts per layer equal to $N-1$, yielding the lowest possible depth $K=N$. The opposite case is illustrated in Fig.~\ref{fig:scheme_variants_App}b, where the phase layers include only single phase shifts, thus, the total depth $K=N^2-2$ is maximal. Finally, yet another scheme variety, depicted in Fig.~\ref{fig:scheme_variants_App}c, permits a number of configurations, in which the number of phases in the phase layers can be arbitrary and can differ from layer to layer. Therefore, in this case, the depth can take values in the range from $N+1$ to $N^2-3$.

\section{Example of a linear-optical CNOT gate}
\label{app::CNOT_gate}

\begin{figure}[h!]
	\includegraphics[width=0.3\linewidth]{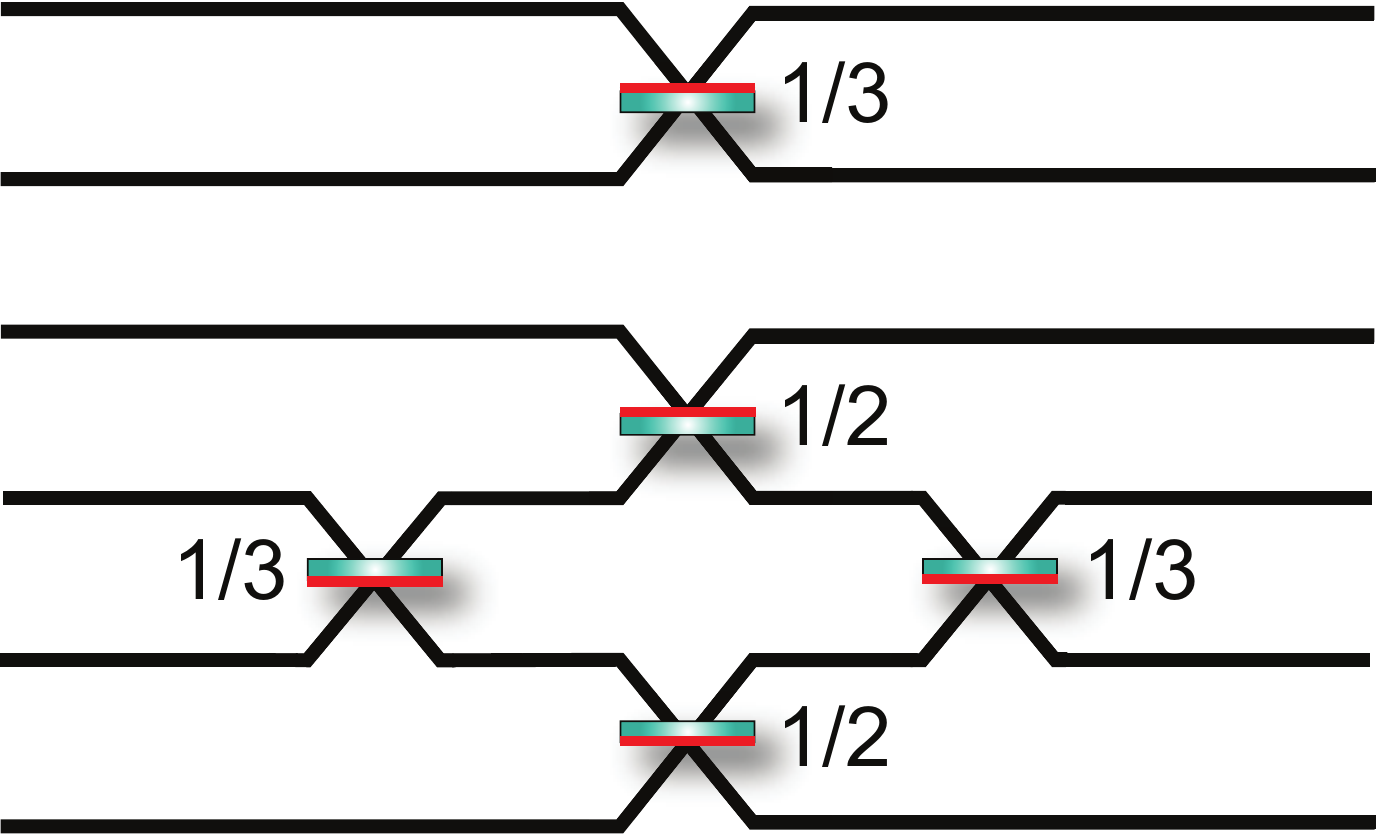}
	\caption{\label{fig:lo_cnot} Optical scheme of a $6$-channel interferometer exploited as an implementation of a linear-optical CNOT gate~\cite{CNOT_Ralph}. The black lines represent channels. The elements on the channel crossings are beam-splitters with specific reflection coefficients indicated on the right. The beam-splitter reflective surface is colored in red to indicate the one where a sign change in reflection occurs.}
\end{figure}

As a specific example of a multi-channel transformation, we consider a $6$-channel interferometer scheme depicted in Fig.~\ref{fig:lo_cnot}, which is exploited in linear-optical quantum computing as an implementation of a CNOT logical gate~\cite{CNOT_Ralph} for dual-rail encoded qubits. Fig.~\ref{fig:lo_cnot} illustrates its canonical representation in the form of a cascaded scheme constructed from standard two-mode beam-splitters. The transfer matrix of the interferometer has the following form:
\begin{equation}\label{eqn:Ucnot}
U_{\text{CNOT}}=\left(
\begin{array}{cccccc}
-\frac{1}{\sqrt{3}} & \sqrt{\frac{2}{3}} & 0 & 0 & 0 & 0 \\
\sqrt{\frac{2}{3}} & \frac{1}{\sqrt{3}} & 0 & 0 & 0 & 0 \\
0 & 0 & \frac{1}{\sqrt{3}} & \frac{1}{\sqrt{3}} & \frac{1}{\sqrt{3}} & 0 \\
0 & 0 & \frac{1}{\sqrt{3}} & -\frac{1}{\sqrt{3}} & 0 & \frac{1}{\sqrt{3}} \\
0 & 0 & \frac{1}{\sqrt{3}} & 0 & -\frac{1}{\sqrt{3}} & -\frac{1}{\sqrt{3}} \\
0 & 0 & 0 & \frac{1}{\sqrt{3}} & -\frac{1}{\sqrt{3}} & \frac{1}{\sqrt{3}} \\
\end{array}
\right).
\end{equation}

To demonstrate that the interferometers of our architecture is capable of implementing the transfer matrix \eqref{eqn:Ucnot}, we selected a single $6$-channel transfer matrix $V$ for all the mixing layers by generating a unitary matrix at random from a Haar-uniform distribution. The elements of this transfer matrix are listed in Table~\ref{tab:static_block}. Then, we found the values of variable phase shifts that reconfigure our interferometer to approximate the transfer matrix (\ref{eqn:Ucnot}) of the CNOT gate as closely as possible. The corresponding phase shifts found in the optimization procedure are listed in Table~\ref{tab:cnot_phase_shifts}. The achieved residual error of the numerical optimization is quantified by the infidelity value, $1-F$, and in this case amounts to $\sim{}3.3\cdot{}10^{-8}$. Since the infidelity is very small, we conclude that the matrix (\ref{eqn:Ucnot}) is reproduced within the numerical precision.

\begin{table}[h!]
	\resizebox{\textwidth}{!}{
		\begin{tabular}{|c|c|c|c|c|c|}
			\hline
			0.32616497+0.169709745i & 
			-0.38944321+0.385222961i & 
			-0.07487433+0.0140016576i & 
			-0.56092302+0.144894222i & 
			-0.09178299+0.182186722i & 
			-0.37672097-0.199464971i\\
			\hline
			-0.59574492+0.341910943i	& 
			-0.06002442+0.0889329420i & 
			0.16555801+0.101417636i & 
			-0.23277056+0.332764918i	& 
			0.50921371-0.219642136i &	
			0.04267885-0.0685604080i \\
			\hline
			-0.11404973-0.0005635556i &	
			0.50625739-0.0702574186i &
			-0.28217237-0.692726708i &
			-0.22904901-0.008906423i &
			0.0634065+0.0378401459i &
			-0.17627259-0.277850848i \\
			\hline
			0.21269847+0.405463432i &	
			-0.01962757+0.030734163i &	
			0.00386132-0.166814050i &
			0.11549117-0.405381161i &
			0.52450465+0.316155727i &
			-0.12388211+0.439441303i \\
			\hline
			-0.19696007+0.129259070i &	
			-0.2829352+0.136684498i &	
			-0.23585629-0.394794001i	&
			-0.07037407+0.253348908i &
			-0.32129528+0.173338317i &
			0.47736737+0.451640055i \\
			\hline
			0.02205175-0.338176828i &	
			-0.57000713+0.0471133316i &	
			-0.12412351-0.377189984i	& 
			0.43404681+0.108197637i &	
			0.33695188-0.153353789i &	
			-0.13322881-0.213157689i \\
			\hline
		\end{tabular}
	}
	\caption{The block transfer matrix $V$ generated at random from a uniform distribution, which is used to implement the $6$-channel transformation \eqref{eqn:Ucnot}.}\label{tab:static_block}
\end{table}
\begin{table}[h!]
	\begin{tabular}{|c|c|c|c|c|c|c|c|}
		\hline
		\multirow{2}{*}{channel ($j$)} & \multicolumn{7}{c}{phase layer ($m$)} \\ 
		\cline{2-8}
		& $1$ & $2$ & $3$ & $4$ & $5$ & $6$ & $7$ \\
		\hline
		1 & 	
		2.51592377 &	
		3.52826283 & 
		0.87421821 &	
		3.24176442 & 
		2.92188993 &	
		1.24223769 & 
		4.30510568 \\
		\hline
		2 & 
		0.96157148	& 
		2.57238693 &	
		4.75174626 &	
		2.81009478 &	
		3.53499635 &	
		3.46774956	& 
		6.21377173 \\
		\hline
		3 & 
		5.5529059 &	
		3.32914678 & 
		1.0145169 & 
		6.25955126	& 
		1.27532946 & 
		0.43181492 &	
		0.91744755\\
		\hline
		4 & 
		1.61215433 &	
		2.30343223 &	
		3.12035203 &	
		4.50728974 &	
		1.78382232 &	
		4.68552694 &	
		2.70720777 \\
		\hline
		5 & 
		1.50025594	&	
		2.0348859	& 
		1.36720772 &	
		4.31057832 &	
		1.88508855	& 
		2.81028466 &	
		0.20069949\\
		\hline
	\end{tabular}
	\caption{\label{tab:cnot_phase_shifts} The values of phase shifts $\varphi_j^{(m)}$ (in radians) that configure the $6$-channel interferometer with the block transfer matrix $V$ shown in Table \ref{tab:static_block} to implement the matrix \eqref{eqn:Ucnot}. The achieved infidelity is $1-F\sim{}3.3\cdot{}10^{-8}$.}
\end{table}

\section{Robustness of the interferometer architecture} \label{app::robustness_interferometer}

Our numerical results suggest that a wide range of unitary matrices, which occupy the major part of the unitary group, can be used as block transfer matrices $V^{(m)}$ for universal interferometers of our architecture. Here, we demonstrate that our architecture is resilient to fabrication errors and constraints, imposed by the technology at hand. 

A model of perturbation which should be used for the block matrices is crucially dependent on a the specifics of the manufacturing facility at hand. As a proof-of-principle, here we model the perturbations by the following procedure. We take the unperturbed matrix for the mixing layer to be the matrix $V_0=V$, which is enlisted in Table~\ref{tab:static_block}. Secondly, we generate a perturbation matrix $R$, which is a Haar-random unitary obtained similarly to $V$. Then, a new parametrized matrix is constructed: $V_\alpha = (1-\alpha)V_0+\alpha R$, with $\alpha$ quantifying the perturbation strength. However, the mixture of two unitary matrices is not unitary, therefore, we then use a singular value decomposition:
\begin{equation}
(1-\alpha)V_0+\alpha R = W_1^{(\alpha)}DW_2^{(\alpha)},
\end{equation}
by means of which the parametrized unitary can be derived as: $V_\alpha = W_1^{(\alpha)}W_2^{(\alpha)}$. In the limiting cases: $V\left(\alpha=0\right)=V_0,V\left(\alpha=1\right)=R$.
We constructed the 6-channel interferometer scheme of our architecture with 6 different blocks, where the first one is $V^{(1)}=V_0$ and the other five having transfer matrices generated independently by the aforementioned procedure with parameter $\alpha=0.1$. By using the numerical optimization algorithm to calculate phase shifts that implement the matrix $U_{\mathrm{CNOT}}$, we arrive at infidelity values $1-F \sim 10^{-8}$ -- the typical error of the optimization algorithm.

\section{Worst-case performance analysis}

Not all transfer matrices may be used for the mixing block. Indeed, at least a unity matrix $I$ and permutations as mixing layers obviously will not allow for good approximation quality. At the same time in our numerical experiments we generated mixing matrices at random and have never encountered a case where the optimization procedure failed to converge. It is therefore interesting to understand, how close to the worst-case layer should one get to have noticeble reduction in the approximation fidelity. Parametrized families of matrices $V^{(m)}_{\alpha}$ allow us to analyze the worst-case performance of our scheme.

In our model for worst-case performance, the block transfer matrices $V^{(m)}_{\alpha}$ have been constructed, that depend on the parameter $\alpha$ taking values from $0$ to $1$. The value of $\alpha=0$ describes an error-free unconstrained scenario when $V^{(m)}_{\alpha=0}$ are known to be suitable for the universal interferometer, while $\alpha=1$ corresponds to the opposite case, when the block transfer matrices are definitely of no use for the interferometer. Namely, we generated unitary matrices $V^{(m)}_0$ at random from a Haar-uniform matrix distribution and independently for each layer, following the procedure from the main text of the paper. Then, these matrices are "corrupted" by the $\alpha$-weighted sum:
\begin{equation}\label{eqn:Talpha}
T^{(m)}_{\alpha}=(1-\alpha)V^{(m)}_0+\alpha{}R,
\end{equation}
where $R$ is a matrix that is known to represent the worst case for static multi-channel block. There exist a number of matrices $R$ that are obviously unsuitable for universal inteferometer, such as diagonal and block-diagonal matrices, which do not provide necessary inter-channel mixing. We  chose $R$ to be an identity matrix $I$, that can be regarded as one of the worst choices for the block transfer matrix (on par with all the diagonal matrices). Since the sum \eqref{eqn:Talpha} is generally not  unitary, the singular value decomposition was used:
\begin{equation}
T^{(m)}_{\alpha}=W_{\alpha}^{(m,L)}D_{\alpha}^{(m)}W_{\alpha}^{(m,R)},
\end{equation}
with which the $\alpha$-parametrized unitary is derived as 
\begin{equation}
V^{(m)}_{\alpha}=W_{\alpha}^{(m,L)}W_{\alpha}^{(m,R)}
\end{equation}
Obviously, $V^{(m)}_{\alpha=0}=V^{(m)}_{0}$ and $V^{(m)}_{\alpha=1}=I$, so by continuously varying the parameter $\alpha$ one can tune between the good choice of the block transfer matrix with which the interferometer is universal and the worst choice, respectively. 
\begin{figure}
	\includegraphics[width=0.4\linewidth]{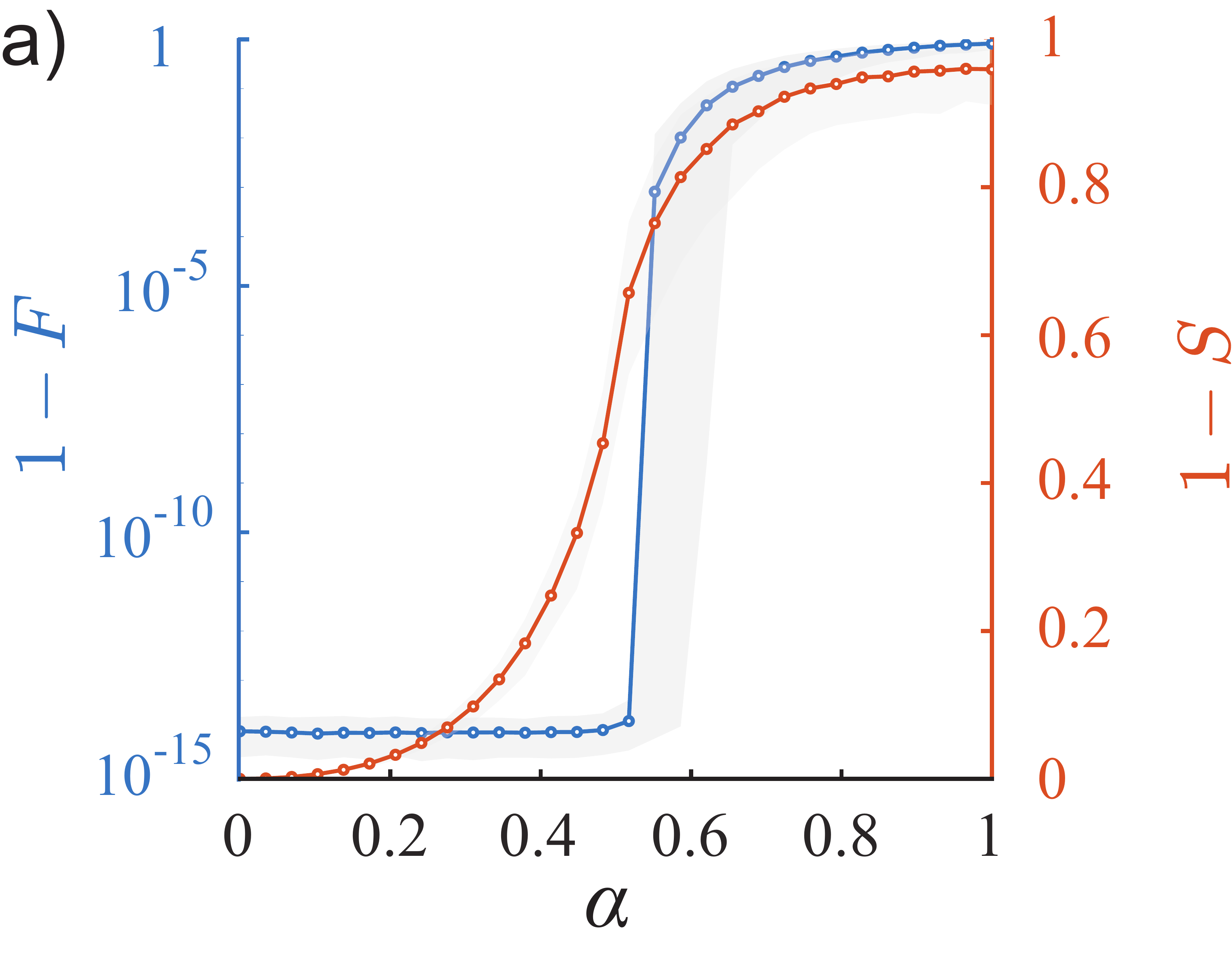}\vspace{0.5cm}
	\includegraphics[width=0.4\linewidth]{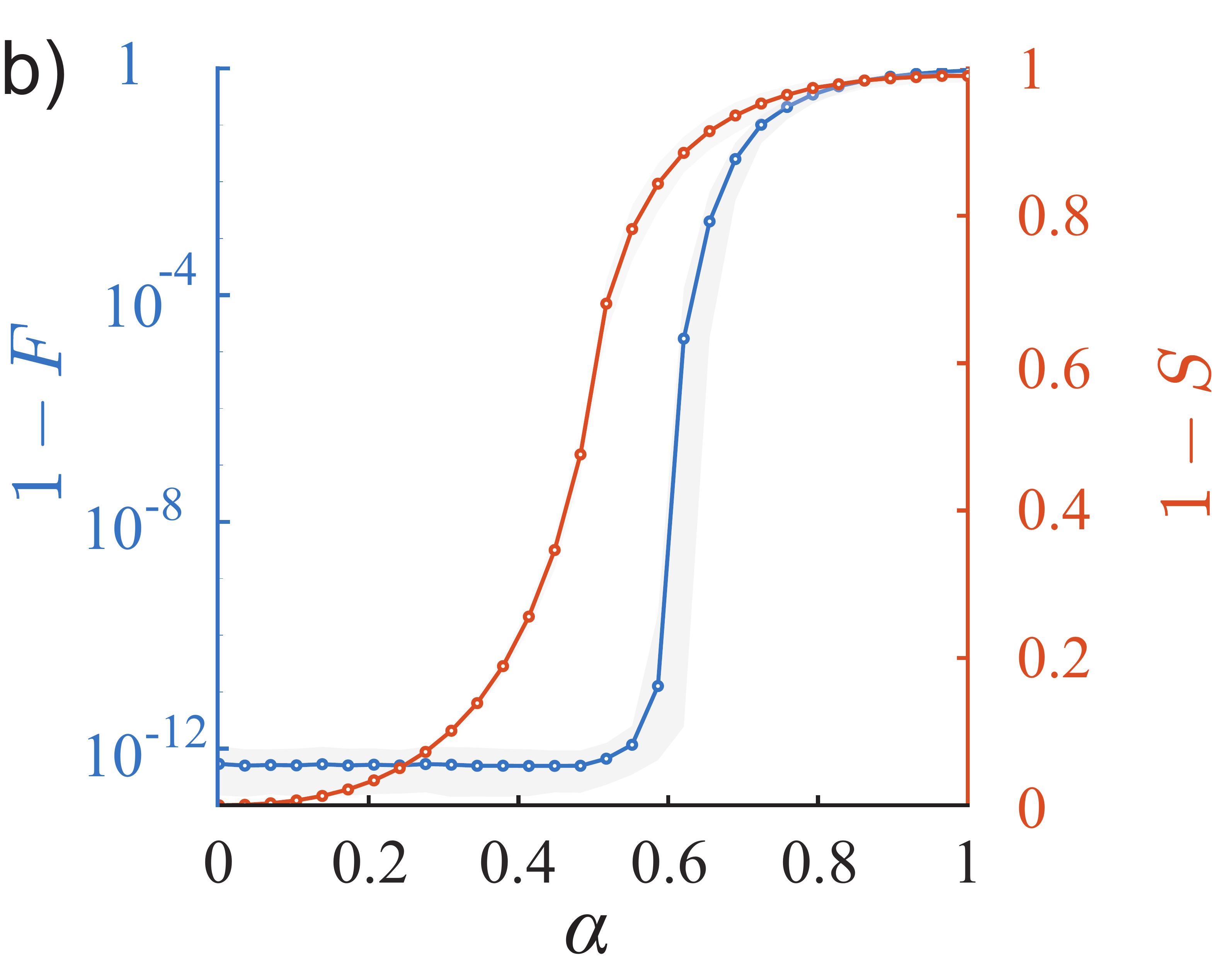}
	\caption{\label{fig:corrupted_blocks} Infidelity of transformation $1-F$ and infidelity of the block transfer matrices $1-S$ as a function of parameter $\alpha$ at $N=5$ (a) and $N=10$ (b). The dependencies were obtained for $300$ randomly generated matrices $U_0$ to be implemented by the interferometer for each value of $\alpha$. For each new matrix $U_0$ and new value of $\alpha$ the block transfer matrices $V^{(m)}_0$ were generated at random. The circles represent average infidelity values; the lower and upper boundaries of the shaded regions are the averages of the $10$ best and worst observed infidelities, respectively.}
\end{figure}
To quantify the similarity between the transfer matrices for a single block we use the fidelity measure averaged over block transfer matrices: 
\begin{equation}
S=\frac{1}{N}\sum_{m=1}^NF(V_0^{(m)},V^{(m)}_{\alpha})=\frac{1}{N^3}\sum_{m=1}^N\left|\text{Tr}(V_0^{(m)\dagger}V^{(m)}_{\alpha})\right|^2.
\end{equation}
%

In order to investigate how the degree of admixture of the "bad" matrices (quantified by $\alpha$) influences the transformation quality, we have performed the numerical analysis. For this purpose, we studied the distribution of fidelities obtained for a set of $300$ randomly generated target unitary matrices $U_0$, which were to be implemented by the interferometer, for each value of $\alpha$. Each random matrix $U_0$ was accompanied by a new set of block transfer matrices $V^{(m)}_{\alpha}$, also generated at random. Fig.~\ref{fig:corrupted_blocks} shows the obtained results where the dependencies of the transformation infidelity $1-F$ and infidelity between the mixing blocks $1-S$ (averaged over the all blocks in the scheme) are plotted. As can be seen from the figure, while infidelity between the block transfer matrices  grows monotonically with respect to $\alpha$, the transformation infidelity exhibits a threshold behavior with the small constant value for $\alpha \lesssim 0.5$ defined by the accuracy of the numerical algorithm used. The threshold where fidelity starts increasing is observed for quite large values of $\alpha$, suggesting additional error resilience of our interferometer architecture.

\section{Computational resources of the numerical optimization} \label{app::optimization_runtime}

Finding the values of tunable phase shifts that make the programmable interferometer to implement the target transformation can demand significant computational resources, especially when optimization is performed on-line with a real interferometer device.  There are numerical methods and software packages currently available that enable effective optimization of multichannel interferometers of previously known architectures (see, e.g. \cite{MillerPhysRevApplied}). Here, to demonstrate that our architecture does not require extraordinary computational power for re-programming, we used the basin-hopping optimization algorithm, which is a part of the SciPy package. 

\begin{figure}
	\includegraphics[width=0.4\linewidth]{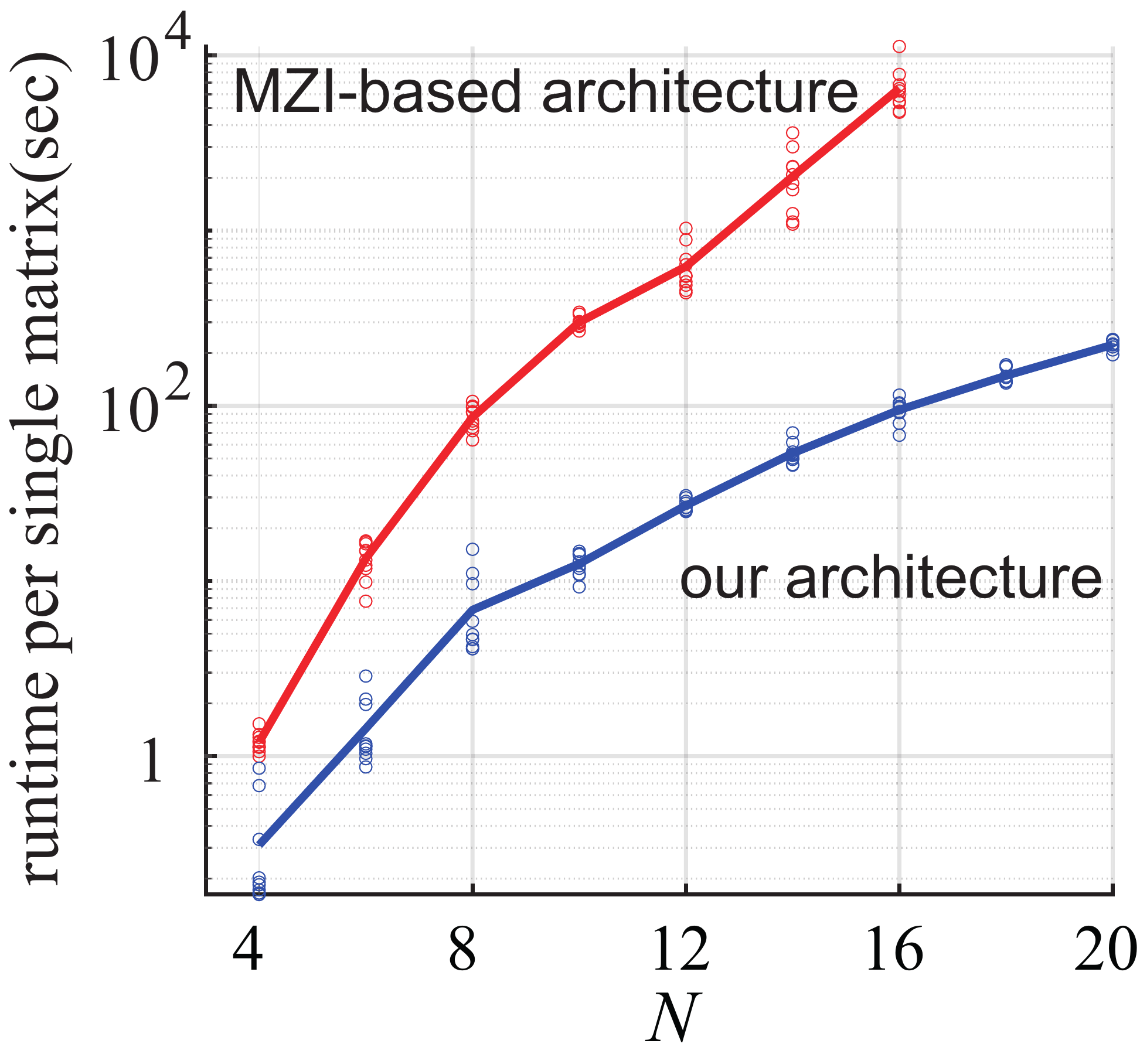}
	\caption{\label{fig:runtime} The  runtime of the basin-hopping optimization algorithm used to find the values of tunable phase shifts as a function of the interferometer size $N$. Random transfer matrices were generated and optimized for each value of $N$ and each architecture. Solid curves correspond to an average over $10$ matrices, while open circles represent the individual data points.}
\end{figure}

We performed numerical optimization for interferometers of different size $N$ and tracked the time it took to obtain high fidelity values satisfying the threshold condition $1-F<10^{-5}$. It should be noted that for the major part of the instances the resultant infidelity was far better than the threshold value. Both interferometers of our architecture and the MZI-based architecture by Clements et al.~\cite{Clements2016} were modelled using the same CPU, thereby providing the possibility to compare the computational costs. Fig.~\ref{fig:runtime} shows the dependence of the runtime as a function of the interferometer size $N$. As can be seen from the figure, the scaling of the optimization runtime with $N$ is much better for the interferometers of our architecture than for the MZI-based counterpart. Notice, however, that a specific interferometer architecture generally requires a specific optimization algorithm, which is optimal for this architecture and can be sub-optimal for another. Devising a specialized optimization algorithm for our architecture is beyond the scope of our work.

\section{The effect of finite phase shift accuracy on the transformation quality} \label{app::finite_accuracy}

In practice, one can not set the values of the tunable phase shifts with arbitrary precision. To study the  negative effect related to phase imprecision on the transformation quality, we have performed numerical analysis and have shown that even moderate precision allows for achieving very-high-fidelity transformations in the interferometers of our architecture. We quantified the phase accuracy by the number of bits $n$ that is guaranteed by the actual phase shifter elements. Thus, the values of the phase shifts are written as $\varphi_j=\varphi_{j0}+\delta\varphi_j$, where $\varphi_{j0}$  describes the $j$th phase in the ideal case,  $\delta\varphi_j$ is the deviation induced by the phase shifter ($j=1\ldots{}N^2-1$). Obviously, in terms of bits, the phase shift is said to provide the $n$-bit precision if $\delta\varphi_j\le{}2\pi/2^{n+1}$. 

Without loss of generality, we consider the specific transfer matrices of the static blocks to be the discrete Fourier transform matrix, i.e. $V^{(m)}=U_{DFT}$  for all $m=1\ldots{}N$. We also consider the performance of the widely-used architecture by Clements et al.~\cite{Clements2016} for comparison. For a given $N$, we obtained the transfer matrix $U_0$ which needs to be implemented by the interferometer by generation of $\varphi_{j0}$ at random from a uniform distribution. Then, the deviations $\delta\varphi_j$ were added to $\varphi_{j0}$  and a new transfer matrix $U$ was calculated accordingly. 

\begin{figure}
	\includegraphics[width=0.35\linewidth]{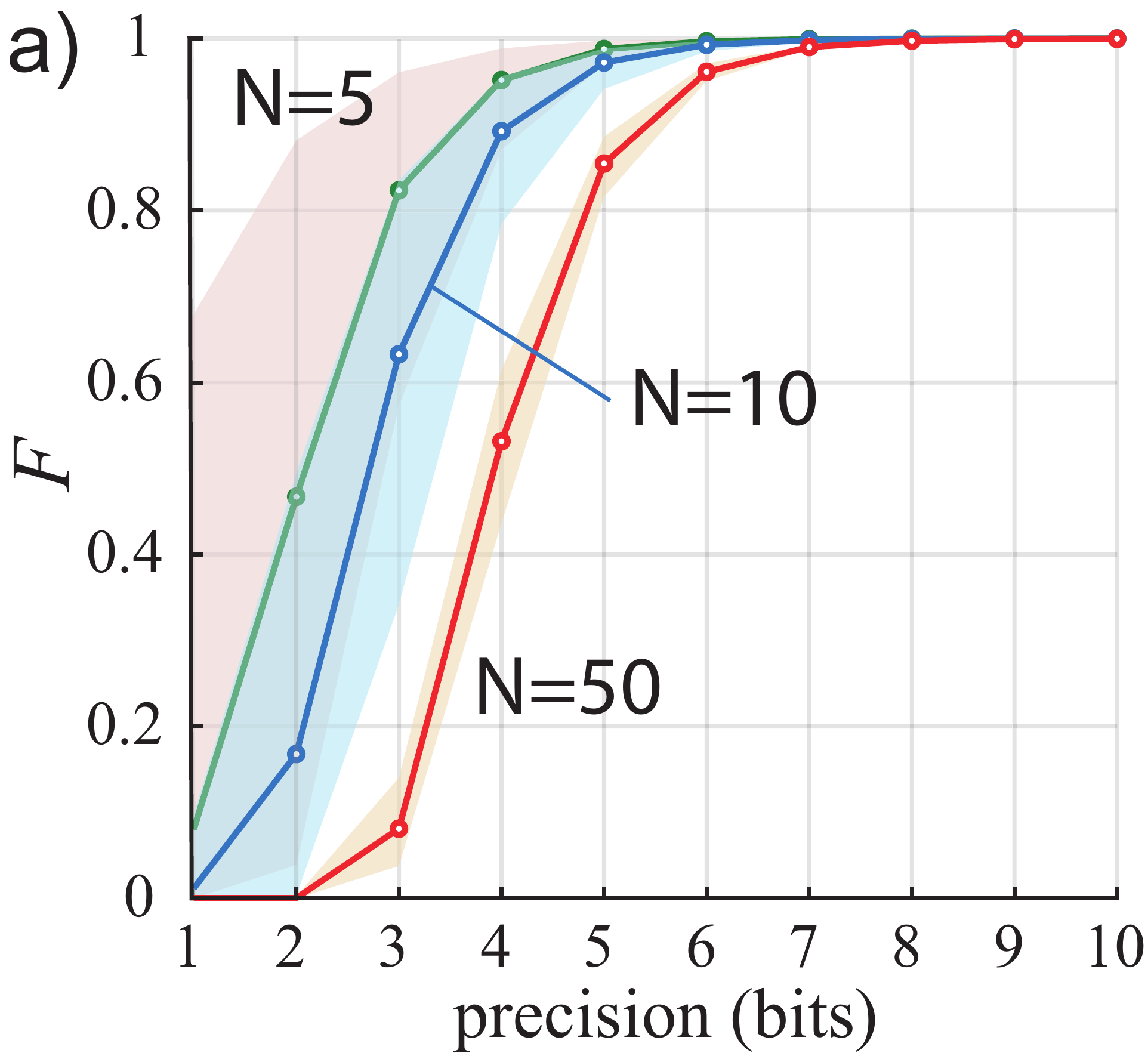}\hspace{0.5cm}
	\includegraphics[width=0.35\linewidth]{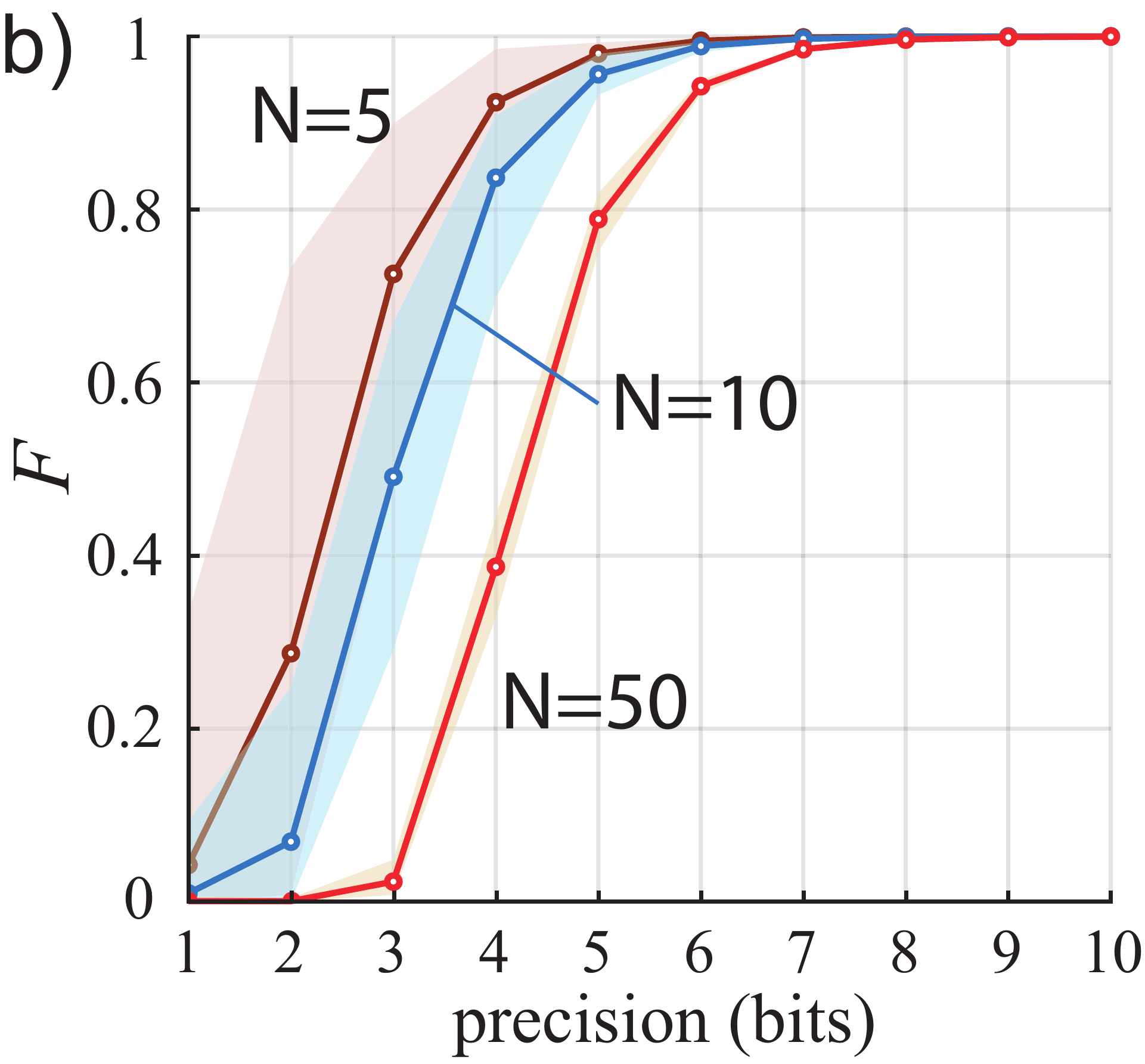}
	\caption{\label{fig:phase_accuracy} The effect of finite precision in setting the tunable phase shifts on the transformation fidelity for interferometers of our architecture (a) and the  architecture proposed in~\cite{Clements2016} (b). To derive these dependencies, a  matrix $U_0$ to be implemented were obtained by generating phase shifts $\varphi_{j0}$ at random and then  for each value of precision bits $n$ $1000$ random phase errors $\delta\varphi_j$ were generated to obtain a series of $1000$ matrices $U$. The circles correspond to average fidelity, while the upper and lower borders of the shaded regions correspond to the worst and best fidelity values, respectively. }
\end{figure}

To have a specific model of inaccuracies $\delta\varphi_j$, the errors $\delta\varphi_j$ were independently drawn from the uniform distribution in the range from $-2\pi/2^{n+1}$ to $2\pi/2^{n+1}$. For each target matrix $U_0$ at each $n$  a series of $1000$ matrices $U$  were generated this way. Fig.~\ref{fig:phase_accuracy} shows the dependence of transformation fidelity $F=|\text{Tr}(U^{\dagger}U_0)|^2/N^2$  as a function of precision bits $n$. As can be seen from the figure, the interferometers of our architectures do not have special requirement on the phase shifts precision, as they are as tolerant as the known counterpart by Clements et al.~\cite{Clements2016}; $10$ bits precision is enough to approximate an $N$-channel unitary matrix with high fidelity for $N\sim{}100$, which is not a problem with the current technologies.

\section{The decomposition structure of  SU(3) transformation}\label{app::transformation_structure}

Let us now develop a theoretical framework describing the capabilities of the unitary decomposition. Firstly we note that if arbitrarily long sequences of fixed transformations are allowed, a set of only two different matrices is enough to generate any $\mathrm{SU}(N)$ transformation \cite{5M}. The proof is straightforward: the first matrix is an irrational phase shift, its powers densely cover all phase shift matrices, and the second matrix is the mixing $U^{(DFT)}$ unitary (4). The question of constructing an arbitrary unitary with a sequence of $U^{(DFT)}$ and phase shifts of a \textit{fixed} length is an open question. Here we provide strong evidence in support of the claim, that it is possible to achieve this goal with the sequence (3).

We have worked out preliminary considerations on the structure of the manifold described by (3).  The tangent space of  transformation $U$ manifold at $N=3$ is 8-dimensional, which means that in principle the transformation is flexible enough to reconstruct the $\mathfrak{su}(3)$ algebra. However, there exist zero-measured submanifolds on which some of the phase factors from the set $\left\{\Phi^{(i)}\right\}$ are linearly dependent in the first order. This means that the coordinates are singular on some submanifolds analogous to the pole in polar or spherical coordinates. Identity of the group lies in the singular submanifold. We show explicitly, that it indeed is a coordinate singularity and the group manifold in the vicinity of the unity matrix is smooth. We do not claim the rigorous proof of universality, but we hope that our considerations may give a deeper insight into the problem.

We study a particular example of tritters (quantum fourier transform matrices) representation of an $\mathrm{SU}(3)$ group, i.e. general 3 by 3 unitary matrices with $\det U=1$ (in what follows matrices $U, \bar U\in SU(3)$). The representation is as follows
\begin{equation}
U= \Phi_{12} T \Phi_{34} T \Phi_{56} T \Phi_{78}\,,
\end{equation}
or one can add an additional tritter matrix
\begin{equation}
\bar U=T U = T \Phi_{12} T \Phi_{34} T \Phi_{56} T \Phi_{78}\,,
\end{equation}
where $$\Phi_{jk}=\left(
\begin{array}{ccc}
\exp(i\varphi_j) & 0 & 0 \\
0 & \exp(i\varphi_k) & 0 \\
0 & 0 & \exp(-i(\varphi_j+\varphi_k)) \\
\end{array}
\right)\,,
$$
with parameters $\varphi_i$, and
$$T=\frac{-i}{\sqrt{3}}\left(
\begin{array}{ccc}
1 & 1 & 1 \\
1 & w & w^2 \\
1 & w^2 & w \\
\end{array}
\right)
,\quad w=\exp\left(i\frac{2\pi}{3}\right)\,,$$

A general element of the $\mathrm{SU}(N)$ group can be represented as an exponent of an algebra generator, in particular, a unitary matrix (fundamental representation) is an exponent of an anti-Hermitian matrix. In the case of phase rotating gates a unitary matrix $\Phi_{jk}$ is an exponent of diagonal Gellmann matrices
\begin{equation}
\Phi_{jk}=\exp\left(-\frac{\varphi_j}{2} \lambda_{7}+\left(\varphi_k+\frac{\varphi_j}{2}\right) \lambda_{8}\right)\,,
\end{equation}
where Gellmann matrices form the basis in the $\mathrm{SU}(3)$ algebra.
Below we use the following notations for Gellmann matrices:
\begin{equation}
\lambda_1=\left(
\begin{array}{ccc}
0 & i & 0 \\
i & 0 & 0 \\
0 & 0 & 0 \\
\end{array}
\right), \quad
\lambda_2=\left(
\begin{array}{ccc}
0 & 0 & i \\
0 & 0 & 0 \\
i & 0 & 0 \\
\end{array}
\right),
\end{equation}
\begin{equation}
\lambda_3=\left(
\begin{array}{ccc}
0 & 0 & 0 \\
0 & 0 & i \\
0 & i & 0 \\
\end{array}
\right), \quad
\lambda_4=\left(
\begin{array}{ccc}
0 & 1 & 0 \\
-1 & 0 & 0 \\
0 & 0 & 0 \\
\end{array}
\right),
\end{equation}
\begin{equation}
\lambda_5=\left(
\begin{array}{ccc}
0 & 0 & 1 \\
0 & 0 & 0 \\
-1 & 0 & 0 \\
\end{array}
\right),\quad
\lambda_6=\left(
\begin{array}{ccc}
0 & 0 & 0 \\
0 & 0 & 1 \\
0 & -1 & 0 \\
\end{array}
\right),
\end{equation}
\begin{equation}
\lambda_7=\left(
\begin{array}{ccc}
-2i & 0 & 0 \\
0 & i & 0 \\
0 & 0 & i \\
\end{array}
\right),\quad
\lambda_8=\left(
\begin{array}{ccc}
0 & 0 & 0 \\
0 & i & 0 \\
0 & 0 & -i \\
\end{array}
\right).
\end{equation}
Diagonal matrices can be permutated with tritter gates due to the following conjugation property
\begin{equation}
\label{l7}
-T \lambda_7 T^{-1} = \lambda_1+\lambda_2+\lambda_3\,,
\end{equation}
\begin{equation}
\label{l8}
-\sqrt{3}~T  \lambda_8 T^{-1} = \lambda_4-\lambda_5+\lambda_6\,.
\end{equation}

Since $T^4=\mathds{1}$, formulae (\ref{l7}) and (\ref{l8}) enables one to express $\bar U$ as follows

\begin{eqnarray}
\label{XPXP}
\nonumber \bar U&=&e^{\varphi_1(\lambda_1+\lambda_2+\lambda_3)+\varphi_2(\lambda_4-\lambda_5+\lambda_6)}~e^{\varphi_3\lambda_7+\varphi_4\lambda_8} \times\\
&\times&e^{\varphi_5(\lambda_1+\lambda_2+\lambda_3)+\varphi_6(\lambda_4-\lambda_5+\lambda_6)}~e^{\varphi_7\lambda_7+\varphi_8\lambda_8}\, ,
\end{eqnarray}
where we have linearly reparametrized $\varphi_i$ for the sake of brevity.

Let us now use the identity which follows from Baker-Campbell-Hausdorff (BCH) formula for the two middle terms in (\ref{XPXP})
\begin{equation}
e^{X}e^{Y}=e^{Y+[X,Y]+\frac{1}{2!}[X,[X,Y]]+\frac{1}{3!}[X,[X,[X,Y]]]+\dots}e^{X}.
\end{equation}
One calculates
	\begin{eqnarray}
	\nonumber &U&=\exp\left[\varphi_1(\lambda_1+\lambda_2+\lambda_3)+\varphi_2(\lambda_4-\lambda_5+\lambda_6)\right]\times\\ \rule{0pt}{0.5cm}
	\nonumber &\times&\exp[\lambda_1(\varphi_5 \cos a-\varphi_6 \sin a)+\lambda_2(\varphi_5 \cos b+\varphi_6 \sin b )+\lambda_3(\varphi_5 \cos (a-b)+\varphi_6 \sin (a-b))+\\
	\nonumber &+&\lambda_4(\varphi_6 \cos a+\varphi_5 \sin a )] 
	-\lambda_5(\varphi_6 \cos b-\varphi_5 \sin b)+\lambda_6(\varphi_6 \cos(a-b)-\varphi_5 \sin(a-b))] \times \\ \rule{0pt}{0.5cm}
	&\times&\exp\left[(\varphi_7+\varphi_3)\lambda_7+(\varphi_8+\varphi_4)\lambda_8\right].
	\label{TPT}
	\end{eqnarray}
The next step is to merge the two first exponents in (\ref{TPT}) by making use of the original BCH formula. The functional independence of coefficients in front of $\lambda_1 - \lambda_6$ in the resulting exponent would conclude the proof of universality. Unfortunately, this calculation appears too involved, and we were not able to sum the full BCH series. 

There is, though, another way to deal with the tritter construction. It simplifies the local (algebraic) study, but still does not allow to prove universality on a group level.
Let us notice that tritter is a matrix exponent of the following combination:
\begin{equation}
T=\exp\left[\frac{\sqrt{3}\pi}{12}\left(\lambda_7-2\lambda_1-2\lambda_2+\lambda_3\right)\right]\, ,
\end{equation}
and it can be conjugated with the phase gate:
	\begin{eqnarray}
	\nonumber A_{1}&=&\Phi(\varphi_1,\varphi_2)\cdot T \Phi^{\dagger}(\varphi_1,\varphi_2)=\\
	\nonumber &=&\exp\left[ \frac{\sqrt{3}\pi}{12}\left(\lambda_7-2\cos(\varphi_1-\varphi_2)\lambda_1+2\sin(\varphi_1-\varphi_2)\lambda_4-2\cos(2\varphi_1+\varphi_2)\lambda_2+2\sin(2\varphi_1+\varphi_2)\lambda_5\right.\right. \\ &+&\left.\left.\cos(\varphi_1+2\varphi_2)\lambda_3-\sin(\varphi_1+2\varphi_2)\lambda_6)\right)\right].
	\end{eqnarray}

This allows one to perform the following procedure:
\begin{eqnarray*}
	&U&=\Phi(\varphi_1,\varphi_2)T\Phi(\varphi_3,\varphi_4)T\Phi(\varphi_5,\varphi_6)T\Phi(\varphi_7,\varphi_8) =
	\\
	&e&^{A_1}\Phi(\varphi_1+\varphi_3,\varphi_2+\varphi_4)T\Phi(\varphi_5,\varphi_6)T\Phi(\varphi_7,\varphi_8) =
	\\
	&e&^{A_1}e^{A_{2}}\Phi(\varphi_1+\varphi_3+\varphi_5,\varphi_2+\varphi_4+\varphi_6)T\Phi(\varphi_7,\varphi_8) =
	\\
	&e&^{A_1}e^{A_{2}}e^{A_{3}}\Phi(\varphi_1+\varphi_3+\varphi_5+\varphi_7,\varphi_2+\varphi_4+\varphi_6+\varphi_8)\, ,
\end{eqnarray*}
where $A_1$ depends only on $\varphi_1$ and $\varphi_2$, $A_{2}$ depends on the sums $\varphi_1+\varphi_3$ and $\varphi_2+\varphi_4$ and so on, hence, all four factors $A_1$, $A_2$, $A_3$ and $P$ are independent.
Now we shift $\varphi_i$ to separate the parameters and simplify the expression:
$$\begin{array}{c}
\varphi_3\rightarrow \varphi_3 -\varphi_1,~~~~~~\varphi_4\rightarrow \varphi_4 -\varphi_2, \\
\varphi_5\rightarrow \varphi_5- \varphi_3 -\varphi_1,~~~~~~\varphi_6\rightarrow \varphi_6- \varphi_4 -\varphi_2, \\
\varphi_7\rightarrow \varphi_7 -\varphi_5- \varphi_3 -\varphi_1,~~~~~~\varphi_8\rightarrow \varphi_8 -\varphi_6- \varphi_4 -\varphi_2,
\end{array}$$
so that each factor depends on it's own couple of $\varphi$'s.
For the sake of brevity we also change the parameters in $A_{i}$ as follows,
$$\begin{array}{c}
a_1=\varphi_1-\varphi_2,~~~~~~b_1=2\varphi_1+\varphi_2, \\
a_2=\varphi_3-\varphi_4,~~~~~~b_2=2\varphi_3+\varphi_4, \\
a_3=\varphi_5-\varphi_6,~~~~~~b_3=2\varphi_5+\varphi_6,
\end{array}$$
so that $A_i$ reads:
	\begin{equation}
	A_i=\frac{\sqrt{3}\pi}{12}\left(\lambda_7-2\cos(a_i)\lambda_1+2\sin(a_i)\lambda_4-2\cos(b_i)\lambda_2+2\sin(b_i)\lambda_5+\cos(a_i-b_i)\lambda_3-\sin(a_i-b_i)\lambda_6\right)\, .
	\end{equation}
In these notations SU(3) element reads
\begin{equation}
U=e^{A_1}e^{A_{2}}e^{A_{3}}\Phi(\varphi_7,\varphi_8)\, .
\label{PTP}
\end{equation}
This form of matrix $U$ may simplify its series decomposition in terms of $\varphi_i$.

\section{Vicinity of a point: an algebraic approach}\label{app::algebraic_approach}

Let us prove that the tangent space to our group manifold in a general point $P$ is 8-dimensional. The tangent space is provided by an equivalency class of a first order expansion of a group element. The corresponding tangent space to identity of a group is a conventional Lie algebra:
\begin{equation}
(e+tA)^{\dagger}(e+tA)=e~~\rightarrow~~A^{\dagger}=-A.
\end{equation}
For a general group element the tangent algebra is twisted by the inverse element:
\begin{equation}
(g+tA)^{\dagger}(g+tA)=e~~\rightarrow~~(g^{\dagger}A)^{\dagger}=-(g^{\dagger}A).
\end{equation}
In other words, element of the algebra is
\begin{equation}
X_{\alpha}=g^{-1}\partial_{\alpha}g
\end{equation}
su(3) commutation relations together with group structure would prove that it is indeed SU(3) group. At least, its universal envelope is SU(3), but finite factor is easy to exclude.
Now we make a first order expansion (for instance using \ref{PTP}):
\begin{eqnarray}
&U&^{-1}\cdot U(a_i+\delta(a_i),\varphi_7+\delta(\varphi_7),\varphi_8+\delta(\varphi_8))=\\
\nonumber &I&+\sum\limits_{i}\delta(a_i)C_i+\delta(\varphi_7)C_7+\delta(\varphi_8)C_8\, .
\end{eqnarray}
The last two terms are obviously linearly independent combinations of $\lambda_7$ and $\lambda_8$. Linear independence of $\lambda_{1-6}$ combinations in $C_i$ was explicitly checked. We do not present the cumbersome answers, since the calculations are totally straightforward. This shows that the tangent space is indeed 8-dimensional.

This result holds in the general point of a group manifold. Unfortunately, there are zero-measured submanifolds where $C_i$ are linearly dependent. Indeed, one can expect that, for instance, if $\varphi_3$ and $\varphi_4$ phases in the $\Phi_2$ gate are zeros, only two out of four parameters in the $\Phi_1$ and $\Phi_3$ gates survive. On the other hand, this does not necessarily make the dimension of the tangent space lower. In full analogy with the pole in polar or spherical coordinates these zero-measured submanifolds may indicate a coordinate singularity. The easiest way to check that the dimension of tangent space did not get lower in the vicinity of degenerate points is to solve numerically the equations of the form
\begin{equation}
U(\varphi_1,..\varphi_8)=U_0\left(1+\epsilon\lambda_i\right)\, ,
\end{equation}
for small $\epsilon$ with an $\epsilon^2$ accuracy. The existence of the solutions for all $\lambda_i$ indicate that the point is regular. The explicit numerical check has been performed around the unity matrix $U_0=I_3$.

However, there is an analytic way to prove that the manifold is regular at a given point $P$. First of all, we construct the metric of the manifold, in case of the Lie group -- the Killing metric, which reads
\begin{equation}
g_{\mu\nu}=Tr(X_{\mu}X_{\nu})=Tr(g^{-1}\partial_{\mu}g g^{-1}\partial_{\nu}g) \, ,
\end{equation}
where $Tr$ stands for trace. The metric is $8$ by $8$ real symmetric matrix. It is degenerate in the points of (coordinate) singularities. But to check whether the manifold itself is singular or not, one has to find Riemann tensor: its regularity would mean regularity of the manifold. This is straightforward differential geometry computation, but very cumbersome. Thus, to simplify the calculation one can use series expansion around point $P$ (one has to keep enough terms of the series due to the second derivatives in expression for curvature and Riemann tensors). We found analytically the Riemann tensor of the manifold in the point corresponding to identity of the group. The tensor is regular, in particular, it is zero, so the manifold is not only regular, but also flat near this point.

\section{Multiport beamsplitter structure} \label{app::multiportBS_structure}

A possible way of physical realization of the proposed structure (3) in linear optics in the case of $N = 3$ is to use three-port beamsplitters, so called \emph{tritters}, as the mode mixing elements $F$, and phase shifters $P$. A tritter is a three-port beamsplitter, which enables three input modes to simultaneously interfere in the three-arm directional coupler, formed by three evanescently coupled waveguides. Experimentally this device can be manufactured, for example, via the femtosecond laser writing technique in a transparent dielectric media \cite{Kowalevicz:05}. The tritter is called balanced when a single photon entering one of three input ports has equal probabilities to exit from each of three output ports. The action of a balanced tritter on can be described by a unitary operator, mapping the input field creation operators $a^{\dagger}$ to the output operators $b^{\dagger}$: $b^{\dagger} = U_{Tr} \ a^{\dagger}$, where $U_{Tr}$ is the matrix of a symmetric balanced tritter.

The matrix of a symmetric balanced tritter can be obtained as follows: let us consider propagation of three photons along the $z$-axis in three equally coupled waveguides. Let $a^{\dagger}_{j}$ be a photon creation operator in a $j$-th waveguide ($j=1,2,3$). According to the coupled mode theory the evolution of the photon creation operator during propagation along the $z$-axis in three equally coupled waveguides could be described by the system \cite{Coupled_mode_theory_Yariv_73}:

\begin{equation}\label{eq::System_T}
\begin{cases}
i\dfrac{da^{\dagger}_{1}}{dz} = \beta a^{\dagger}_{1} + c a^{\dagger}_{2} + c a^{\dagger}_{3}, \\
i\dfrac{da^{\dagger}_{2}}{dz} = c a^{\dagger}_{1} + \beta a^{\dagger}_{2} + c a^{\dagger}_{3},  \\
i\dfrac{da^{\dagger}_{3}}{dz} = c a^{\dagger}_{1} + c a^{\dagger}_{2} + \beta a^{\dagger}_{3}.
\end{cases}
\end{equation}

where $\beta$ is the propagation constant, $c$ -- the coupling coefficient between the neighbouring waveguides ( if $c$ is the same for all pairs of waveguides the tritter is \emph{symmetrical}.)

The system (\ref{eq::System_T}) can be explicitly solved:
\begin{equation}\label{System_T_Sol}
\begin{pmatrix} A^{\dagger}_{1}(z) \\ A^{\dagger}_{2}(z) \\ A^{\dagger}_{3}(z) \end{pmatrix} = U \begin{pmatrix} A^{\dagger}_{1}(0) \\ A^{\dagger}_{2}(0) \\ A^{\dagger}_{3}(0) \end{pmatrix},
\end{equation}
where 
\begin{equation}\label{T_Matrix1}
U = \frac{1}{3} e^{-i\beta z} \begin{pmatrix} d & b & b  \\ b & d & b \\ b & b & d \end{pmatrix},
\end{equation}

\begin{align}\label{T_Matrix2}
&d = \exp(-i2kz) + 2\exp(ikz), \\
&b = \exp(-i2kz) - \exp(ikz).
\end{align}

It can be shown that the interaction length $z = L$ should be chosen as $kL = 2\pi/9$ for a balanced tritter. Then the matrix of a symmetrical balanced tritter is (up to a phase shifts inserted in the input or in the output modes):

\begin{equation}\label{eq::T_Matrix5}
U_{Tr} =U^{(DFT)}_3 = \frac{1}{\sqrt{3}} \begin{pmatrix} 1 & 1 & 1  \\ 1 & e^{i2\pi/3} & e^{i4\pi/3} \\ 1 & e^{i4\pi/3} & e^{i2\pi/3} \end{pmatrix} ,
\end{equation}
which appears to be exactly the Fourier transform matrix (4) in case $N = 3$. 
Here, the subscript index marks the matrix size.

Similarly, it is possible to construct a symmetrical balanced four-path beamsplitter -- a \emph{quarter}. 
Such devices may also be implemented with the aid of femtosecond laser writing techninque \cite{Spagnolo2012}. The quarter matrix is obtained in the same way as that for a tritter, and it is essentially the $4 \times 4$  Fourier transform matrix:
\begin{equation}\label{Q_Matrix}
U^{(DFT)}_4 = \frac{1}{2} \begin{pmatrix} 1 & 1 & 1 & 1 \\ 1 & i & -1 & -i \\ 1 & -1 & 1 & -1 \\ 1 & -i & -1 & i \end{pmatrix}
\end{equation}

Using interference schemes (3) based on these particular multipath beam splitters and phase shifters one may implement an arbitrary unitary transformation in the corresponding dimension.

\end{document}